\documentclass[journal=jpccck,manuscript=article]{achemso}

\usepackage{chemformula} 
\usepackage[T1]{fontenc} 

\usepackage[version=3]{mhchem}
\usepackage{float}
\usepackage{lscape}
\renewcommand\vec\mathbf

\def\blacket<#1>{\langle #1 \rangle}

\author{Kaito Miyamoto}
\email{kaito@mosk.tytlabs.co.jp}
\phone{+81-561-71-7973}
\fax{+81-561-63-6279}
\author{Ryoji Asahi}
\affiliation[Toyota Central R\&D Labs., Inc.]
{Toyota Central R\&D Labs., Inc., 41-1, Yokomichi, Nagakute, Aichi 480-1192, Japan}

\title[An \textsf{achemso} demo]
  {Water Facilitated Electrochemical Reduction of \ce{CO2} on Cobalt-Porphyrin Catalysts}

\begin{document}

\begin{abstract}
Cobalt-porphyrin catalyzed reductive decomposition of \ce{CO2} to CO is investigated based on the Koper's water facilitated \ce{CO2} reduction mechanism using simple but accurate protocol based on thermodynamics.
In our protocol, accurate predictions of standard redox potentials and free energy differences are achieved by combining strengths of both density functional theory and experimental observations.
With the proposed protocol, we found that the proton transfer from \ce{H2O} takes place at $-$0.80 V \textit{vs.} RHE at pH=3 through a concerted pathway and, as a result, the key intermediate for the CO generation, i.e., \ce{[CoP-COOH]-} is formed. 
Since the redox potential of the proton transfer agrees well with experimentally observed \ce{CO2} reduction potential, we successfully clarified that \ce{H2O} plays an important role in the reductive decomposition of \ce{CO2} to CO.
This result is valuable not only for understanding the cobalt-porphyrin catalyzed reductive decomposition of \ce{CO2} but also as a guide for the development of new catalysts.
\end{abstract}

\section{\label{sec:intro}INTRODUCTION}
Electrochemical fixation of carbon dioxide is one of the promising and imperative countermeasures to mitigate global warming and energy storage problems, where \ce{CO2} gas is electrochemically converted to fuels and commodity chemicals.~\cite{finn2012, qiao2014, gattrell2006}
The key step for this conversion is the activation of thermodynamically stable \ce{CO2}. 
Especially, the conversion to \ce{CO} is known to be the key and rate determining step to obtain useful chemicals~\cite{gattrell2006, costentin2012, lu2014, liu2016, rosen2011, lin2015} and effective catalysts have been pursued to reduce its large overpotential.~\cite{rosen2011, costentin2012, lu2014, qiao2014, lin2015, shen2015, varela2015, kornienko2015, liu2016, tripkovic2013, cheng2015tailoring, wannakao2017} 
Among such catalysts, cobalt porphyrin complexes are considered to be one of the promising candidates since it selectively reduces \ce{CO2} to \ce{CO} with relatively small overpotential in aqueous solutions.~\cite{sonoyama1999electrochemical, magdesieva2002electrochemical} 
Recently, drastic improvements in terms of stability in aqueous medium, high selectivity in \ce{CO2} reduction, and low overpotential, have been achieved by the immobilization to  the graphite electrode~\cite{atoguchi1991, yoshida1995, tanaka1997, shen2015, hu2017enhanced} or by using as the building blocks of COF (covalent organic frameworks)~\cite{lin2015} and MOF (metal organic frameworks).~\cite{kornienko2015}  
 
To achieve further improvement, understanding of catalytic effects of cobalt porphyrin complexes is quite important and, therefore, analyses using electronic structure calculations have been carried out.~\cite{nielsen2010, leung2010, kortlever2015, shen2016, gottle2017} 
The seminal work in this field was done by Leung et al. \cite{nielsen2010, leung2010}
They investigated the catalytic effect of a cobalt porphine (CoP) molecule on reductive decomposition of \ce{CO2} in water using the combination of quantum chemistry calculations and \textit{ab initio} molecular dynamics simulations. 
Their proposed mechanism starts from the one-electron reduction of CoP (\ce{CoP} + \ce{e-} $\rightarrow$ \ce{CoP-}).
After the binding of \ce{CO2} to the reduced cobalt atom, further reduction takes place ([CoP-\ce{CO2]-} + \ce{e-} $\rightarrow$ \ce{[CoP-CO2]}$^{2-}$).
Then, after the protonation, detachment of \ce{OH-} moiety occurs ( \ce{[CoP-COOH]-} $\rightarrow$ CoP-CO + \ce{OH^-}). 
Their mechanism successfully explained the reaction at pH$=$7 in water. 
However, it is also known that \ce{CO2} reduction takes place in aqueous medium at much lower pH (pH $=$ 3),~\cite{shen2015} which cannot be explained by their mechanism.

Very recently, Yao et al. proposed an interesting reaction mechanism that starts from the \ce{H+} binding to the N site of CoP ([CoP$\cdot$H])$^+$.~\cite{yao2017cobalt}
In their mechanism, \ce{CO2} binds to Co site after the two-electron reduction of [CoP$\cdot$H]$^+$ at around $-$1.23 V vs. SHE.
However, it is questionable that the \ce{H+} always exists on the N site during the operation since it is consumed to reduce \ce{CO2} to \ce{CO} and the binding between the \ce{Co} and \ce{H+} is much stronger (by 0.7 eV) than that of the \ce{N-H+} bond in \ce{CoP-} (Table S1 in the Supporting Information).
Maybe the other main reaction routes also exist.   

All the above mechanisms assume that \ce{H+} plays an important role in the \ce{CO2} reduction.
Contrary to this, Koper et al. proposed a novel \ce{CO2} reduction mechanism in aqueous medium based on their experimental~\cite{shen2015} and computational~\cite{gottle2017} analyses, where \ce{H2O} plays a central role in the \ce{CO2} reduction. 
Their reaction mechanism starts from the \ce{CO2} binding to the negatively charged cobalt-porphyrin complex `M' i.e.,  
\begin{equation}
  \ce{M} + \ce{CO2} + \ce{e-} \rightarrow \ce{[M-CO2]-} \;.
  \label{eq:copco2}
\end{equation}
Then, a proton is transferred from the water molecule and a carboxyl group is created as
\begin{equation}
  \ce{[M-CO2]-} + \ce{H2O} + \ce{e-} \rightarrow \ce{[M-COOH]-} + \ce{OH-} \;.
  \label{eq:H2OOH}
\end{equation}
Finally, CO gas is released by following the similar reaction proposed by Leung et al.~\cite{nielsen2010, leung2010} that
\begin{equation}
  \ce{[M-COOH]-} \rightarrow \ce{M} +\ce{CO} + \ce{OH-} \;.
  \label{eq:reaction3}
\end{equation}
The key step in this mechanism is Eq.~\ref{eq:H2OOH}, where the water molecule not the proton becomes the proton source to facilitate \ce{CO2} reduction.
However, by considering the pH of the pure water which corresponds to the proton concentration of $10^{-7}$ mol/l, it is not so easy to imagine that the water molecule has an ability to provide enough protons so that a large amount of \ce{CO2} reduction takes place.
Therefore further investigations are necessary. 
The goal of this paper is to shed light on the details of the Koper's water facilitated \ce{CO2} reduction mechanism, i.e., Eqs.~\ref{eq:copco2} - \ref{eq:reaction3},~\cite{shen2015, gottle2017} using simple but accurate protocol based on thermodynamics, which combines density functional theory (DFT) and experimental data.

\section{COMPUTATIONAL DETAILS}
\begin{figure}[h]
\centering
  \includegraphics[scale=0.50]{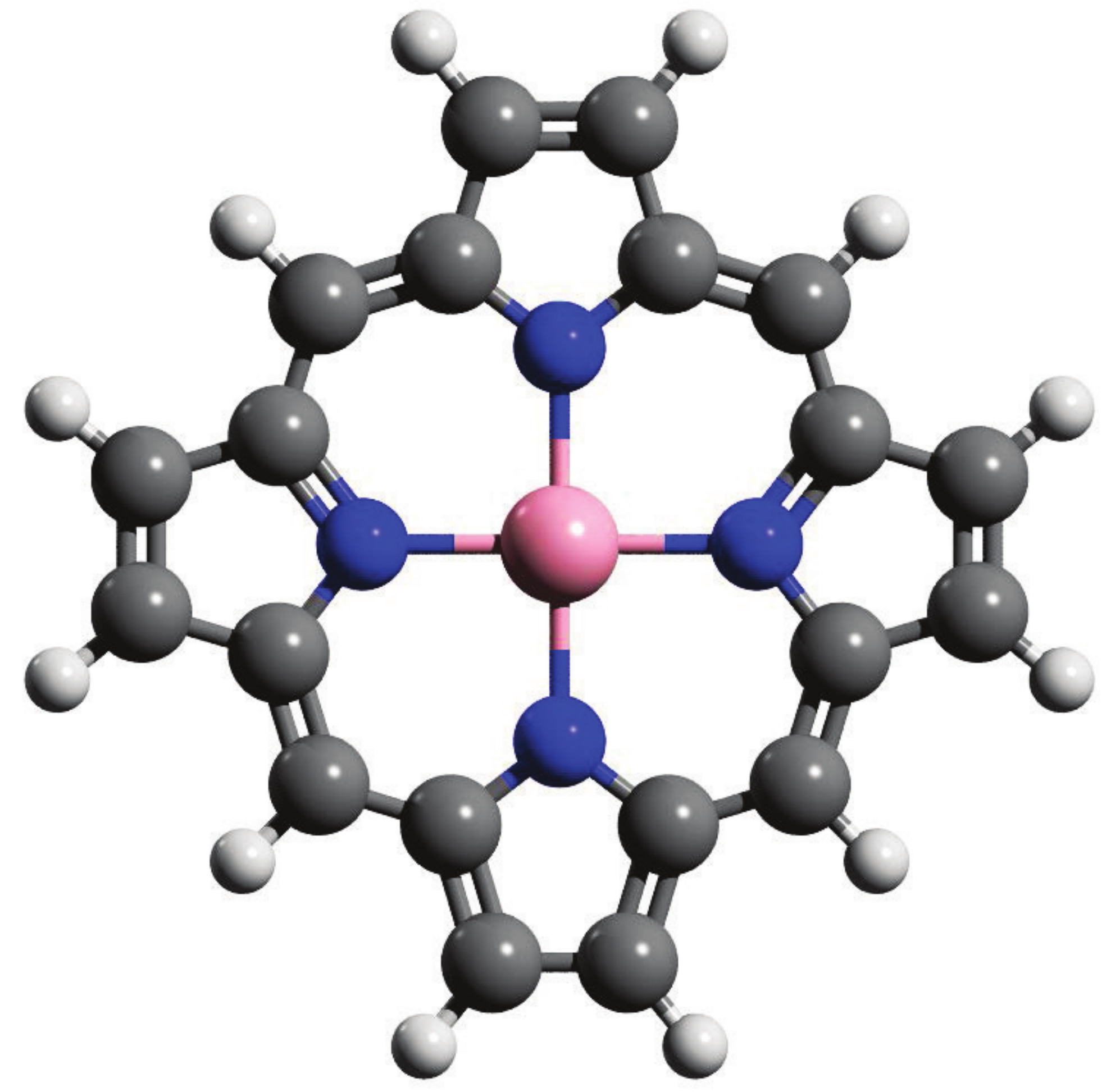}
  \caption{
             The geometry of cobalt porphine(CoP).
             Pink, blue, gray, and white spheres represent Co, N, C, and H atoms, respectively.
             }
  \label{geom}
\end{figure}

In this study, the cobalt porphine (CoP) complex (Fig.~\ref{geom}) is used as the model for the \ce{CO2} reduction catalyst, i.e., M=CoP in Eqs.~\ref{eq:copco2} -- \ref{eq:reaction3}.
Gibbs free energies of metal complexes, such as CoP, \ce{[CoP-CO2]-}, and \ce{[CoP-COOH]-}, are evaluated using DFT where  B3P86~\cite{b3lyp, perdew1986} and 6-311++G**~\cite{krishnan1980, mclean1980, clark1983} are used as the exchange-correlation functional and basis set respectively. 
B3P86 is known to reproduce the geometries of cobalt complexes.~\cite{solis2011theoretical}
The solvation effect is taken into account using the polarizable continuum model.
SMD model~\cite{marenich2009} is selected as the continuum model and default water parameters in Gaussian 09~\cite{g09} are employed. 
The validity of the method was confirmed using geometries and one-electron reduction potentials of cobalt tetraphenylporphyrin (CoTPP) and cobalt phthalocyanine (CoPc). 
As shown in Table S2 and S3 in the Supporting Information, the errors in the geometries are less than 0.02 {\AA}  and that in the reduction potentials are less than 0.2 V, meaning the results well agree with experimental data. 

The spin state of each system is set to be a low-spin state by following the previous reports.~\cite{nielsen2010, shen2016}
All the geometries were confirmed to be minima by carrying out frequency calculations.
We checked the stability of the wave functions for all the species and confirmed that all the wave functions except for \ce{CoP-} are stable.
As for \ce{CoP-}, we carried out energy and geometry optimization calculations using the broken symmetry wave functions.~\cite{seeger1977, bauernschmitt1996, cramer2013essentials, cramer1998bergman, debbert2000}
The thermal corrections to the Gibbs free energy are computed at 298.15 K.
All the calculations were performed with Gaussian 09.~\cite{g09}

The standard one-electron redox potential relative to the reversible hydrogen electrode (RHE) is computed by
\begin{align}
  E^\circ_\text{RHE} = - \frac{\Delta G^\circ}{F} - E^\text{toSHE} + 0.0592 \times \text{pH }- E^\text{corr}\;,
  \label{eq:e0}
\end{align}
where $F$ and $\Delta G^\circ$ are the Faraday constant and the standard Gibbs free energy difference of the target reaction, respectively; $E^\text{toSHE}$ is the factor to convert reference system from the vacuum level to the standard hydrogen electrode (SHE). We employ the value of 4.28 V\cite{marenich2014, isegawa2016, okoshi2015theoretical} for $E^\text{toSHE}$.
The third term of the right hand side is to convert the reference from SHE to RHE.
The last term ($E^\text{corr}$) is the parameter to correct the error derived from the computational methods. 
The effectiveness of this correction is shown in Refs.~\citenum{solis2011theoretical}, \citenum{solis2011substituent}, and \citenum{huo2016breaking}. 
Since the errors in the one-electron reduction potentials of CoPc and CoTPP using the present computational condition are $-$0.17 and $-$0.19 V respectively, as shown in Table S3 in the Supporting Information, we employ the average value ($-$0.18 V) as $E^\text{corr}$.
Here, it is again worth emphasizing that the errors in the one-electron reduction potentials of CoPc and CoTPP (and hence $E^\text{corr}$) are small.
In this study, we set pH to 3 since comparable experimental data are available.~\cite{shen2015}

In order to obtain the standard redox potential or free energy difference in Eq.~\ref{eq:H2OOH}, 
not only the Gibbs free energies of metal complexes but also an accurate value of a free energy difference between \ce{H2O} and \ce{OH-}, i.e., $\Delta G^\circ_\text{H2O} = G^\circ(\ce{OH-}) - G^\circ(\ce{H2O})$ is necessary. 
This value is obtainable from the experimental p$K_\text{a}$ value of water and the free energy of a single proton in water~\cite{casasnovas2014theoretical} as 
\begin{equation}
  \Delta G^\circ_\text{H2O} = 2.303 R T\text{p}K_\text{a}\text{(\ce{H2O})} 
                             - (G_\text{gas} (\ce{H+}) + \Delta G_\text{solv} (\ce{H+}) 
                                + RT \ln 24.46) \;,
\end{equation} 
where $R$, $T$, and p$K_\text{a}$(\ce{H2O}) denote the gas constant, temperature, and p$K_\text{a}$ value of water (= 14),~\cite{crc} and where $G_\text{gas}(\ce{H+})$ (=$-$6.28 kcal/mol)~\cite{casasnovas2014theoretical} and $\Delta G_\text{solv} (\ce{H+})$ (= $-$265.9 kcal/mol)~\cite{kelly2006aqueous} are the free energy of a proton in gas phase and the solvation free energy of a single proton in water respectively. 
If $T$ is set to be 298.15 K, $\Delta G^\circ_\text{H2O}$ becomes 12.549 eV.

\section{RESULTS AND DISCUSSION}
\begin{figure}[h]
\centering
  \includegraphics[scale=0.50]{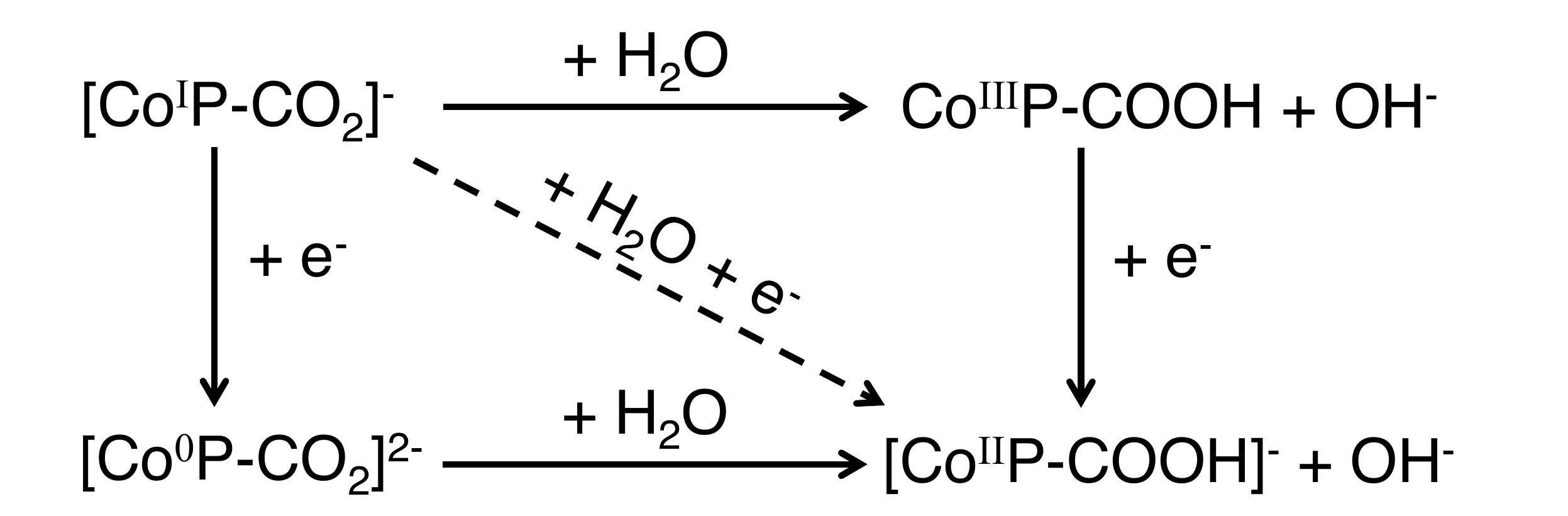}
  \caption{
             Possible reaction routes starting from \ce{[CoP-CO2]-}.
             The solid lines represent the sequential pathways and the dashed line the concerted pathway.
             }
  \label{SandC}
\end{figure}

In this paper, both sequential and concerted reaction mechanisms, shown in Fig.~\ref{SandC}, are investigated for the key step (Eq.~\ref{eq:H2OOH}) with M = CoP, that take place after the formation of \ce{[CoP-CO2]-} (Eq.~\ref{eq:copco2}). 
While the comparisons between the sequential and concerted reaction pathways were made in detail for the proton-coupled electron transfer in the electrochemical reduction of \ce{CO2} using cobalt-porphyrin catalysts,~\cite{gottle2017, koper2013theory}
the present study is different from the previous report in that not the \ce{H+} in the aqueous medium but the \ce{H2O} molecule becomes the proton source to facilitate the \ce{CO2} reduction. 
Here, it is noteworthy that the \ce{[CoP-CO2]-} formation reaction (Eq.~\ref{eq:copco2}) takes place at $-$0.43 V \textit{vs.} RHE under the condition that pH = 3 using the present computational condition, which is close to the one-electron reduction potential of CoP ($-$0.43 V). 

At first, we discuss the sequential reactions. In this mechanism, there are two possible reaction routes, i.e., \ce{[CoP-CO2]-} firstly reacts with \ce{H2O} and, then, \ce{e-} comes or vise versa. 
In case we assume \ce{[CoP-CO2]-} reacts with \ce{H2O} first, the reaction route in Eq.~\ref{eq:H2OOH} is decomposed into two elementary reactions as
\begin{equation}
  \ce{[CoP-CO2]-} + \ce{H2O} \rightarrow \ce{CoP-COOH} + \ce{OH-}
  \label{eq:h2oreac}
\end{equation}  
and  the subsequent \ce{e-} transfer reaction given by
\begin{equation}
  \ce{CoP-COOH} + \ce{e-} \rightarrow \ce{[CoP-COOH]-} \;.
  \label{eq:copcoohred}
\end{equation}

The free energy difference of the proton donation reaction (Eq.~\ref{eq:h2oreac}) is shown in Table.~\ref{tab:hbye}. 
The proton donation reaction is the endergonic reaction and the $\Delta G^\circ = 0.50$ eV is quite large.
Therefore, \ce{CO2} reduction does not proceed by this route. 

Next, the reaction route starting from the electron transfer is considered.
Since the redox potential of the reaction: \ce{[CoP-CO2]-} + \ce{e-} $\rightarrow$ \ce{[CoP-CO2]}$^{2-}$ is $-$0.47 V \textit{vs.} RHE (pH = 3), 
this reaction occurs at slightly lower potential than the one-electron reduction potential of CoP ($-$0.43 V, see Table~\ref{tab:cpet2}).
However, the subsequent reaction is endergonic as shown in Table~\ref{tab:hbye}. 
Therefore, \ce{CO2} reduction does not proceed by this route. 
Thus, the water facilitated \ce{CO2} reduction does not proceed by the sequential reactions.
If enough amount of \ce{H+} exists in the electrolyte, \ce{CO2} reduction proceeds via the route: \ce{[CoP-CO2]}$^{2-}$ + \ce{H+}$\rightarrow$ \ce{[CoP-COOH]-} and 
competes with the direct proton reduction (the \ce{H+} reduction to \ce{H2}).
However, by considering the experimental results in Ref.~\citenum{shen2015}, which reports that the direct proton reduction is diffusion limitation at pH = 3, \ce{CO2} reduction via this route may not be dominant.

\begin{table}[h]
  \caption{\ Free energy difference of proton donation reactions from \ce{H2O}.}
  \label{tab:hbye}
  \begin{center}
  \begin{tabular}{lc}
    \hline
    reaction                                                                    & $\Delta G^\circ$ (eV)  \\
    \hline
     \ce{[CoP-CO2]-}       + \ce{H2O} $\rightarrow$ \ce{CoP-COOH}    + \ce{OH-} & 0.50 \\ 
     \ce{[CoP-CO2]}$^{2-}$ + \ce{H2O} $\rightarrow$ \ce{[CoP-COOH]-} + \ce{OH-} & 0.19 \\ 
%
    \hline
  \end{tabular}
  \end{center}
\end{table}

Next, the possibility of the concerted reaction is investigated, where a water molecule and an electron react with \ce{[CoP-CO2]-} in a single step as shown in Fig.~\ref{SandC}.
The standard equilibrium potentials of the concerted reaction as well as the \ce{CoP-}/CoP redox system relative to RHE (pH = 3) are summarized in Table~\ref{tab:cpet2}.
It is noteworthy that the redox potential of the \ce{CoP-}/CoP redox couple is within the range of the one-electron reduction potentials of typical cobalt porphyrin complexes ($-$0.19 to $-$0.44 V relative to RHE at pH = 3).~\cite{felton1966polarographic, behar1998cobalt, shen2015} 

Using our protocol, $E^\circ$ of the concerted reaction is computed to be $-$0.67 V.  
From the experimental analysis,~\cite{shen2015} \ce{CO2} reduction takes place at around $-$0.6 V \textit{vs.} RHE when pH=3, which well agrees with the present result.
This agreement suggests that, at pH $\geq$ 3, the proton supply from \ce{H2O} takes place through the concerted reaction:
\begin{equation}
  \ce{[CoP-CO2]-} + \ce{H2O} + \ce{e-} \rightarrow \ce{[CoP-COOH]-} + \ce{OH-} \;.
  \label{eq:concerted}
\end{equation}

Finally, we investigate the effect of the hydrogen bonds on the standard equilibrium potential of the concerted reaction.
It is reported that both CoP and \ce{CoP-} do not form the hydrogen bond with surrounded water molecules while \ce{[CoP-CO2]-} and \ce{[CoP-COOH]-} forms such bonds at \ce{CO2} or \ce{COOH} moiety.~\cite{leung2010} 
Since $E^\text{corr}$ in Eq.~\ref{eq:e0} does not correct errors derived from the hydrogen bonds in $E^\circ$, we estimate such errors by adding water molecules explicitly.
This method is the so-called cluster-continuum method.~\cite{pliego2002theoretical}

\begin{figure}[h]
\centering
  \includegraphics[scale=0.50]{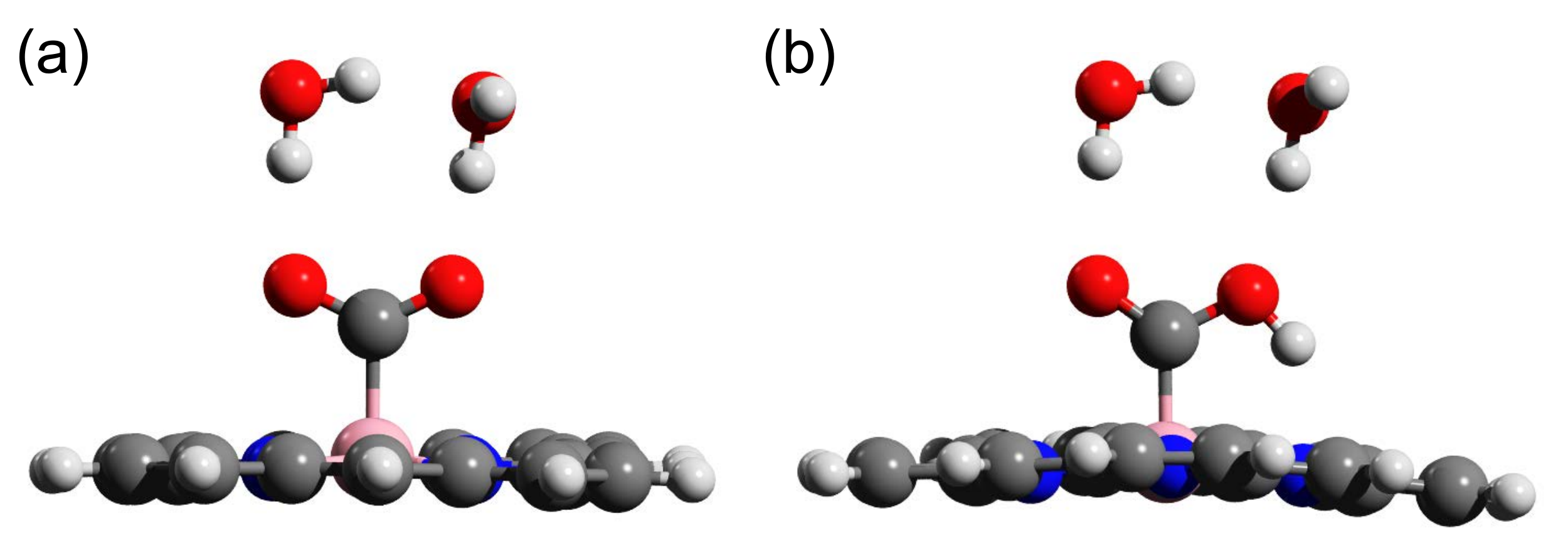}
  \caption{
             The geometries of (a) \ce{[CoP-CO2]-} and (b) \ce{[CoP-COOH]-} with hydrogen-bonded two water molecules.
             Pink, blue, gray, red, and white spheres represent Co, N, C, O, and H atoms, respectively.
             }
  \label{geom2}
\end{figure}

We added two explicit water molecules in order to investigate the effect of the hydrogen bonds to the standard equilibrium potential of the concerted reaction. 
Positions of the two water molecules are determined based on the previous report~\cite{gottle2017} and the detailed geometries are shown in Fig.~\ref{geom2}. 
The number of explicit water molecules is determined based on the previous study~\cite{pliego2002theoretical} which reports that the cluster-continuum method provides p$K_\text{a}$ values of carboxylic acids with the error of around 0.4, corresponding to the energy error of 0.02 eV, if two explicit water molecules are added to the system. 
It is also reported that addition of the two water molecules to the system improves the prediction of the p$K$a values of \ce{CoP-COOH} and \ce{[CoP-COOH]-}. \cite{gottle2017}
Indeed, the computed p$K$a values of \ce{CoP-COOH} and \ce{[CoP-COOH]-} well agree with previously reported values,~\cite{gottle2017, leung2010} i.e., the average error of less than 0.4 as shown in Table S4 in the Supporting Information. This agreement is interesting by considering the fact that the p$K$a values are obtained using completely different methods, i.e., \textit{ab initio} molecular dynamics simulations,\cite{leung2010} the method based on the isodesmic proton-exchange reaction scheme,~\cite{gottle2017} and our protocol. 

The standard equilibrium potential of the concerted reaction using the cluster-continuum model is shown in Table~\ref{tab:cpet2}. 
Although the explicit treatment of the hydrogen bonds slightly lowers the redox potential of the concerted reaction ($-$0.80 V), it still agrees well with experimental \ce{CO2} reduction potentials (around $-$0.60 V).
Therefore, we conclude that, at low pH region (pH of around 3), water-facilitated \ce{CO2} reduction takes place through concerted reaction (Eq.~\ref{eq:concerted}).
                  
\begin{table}[h]
  \caption{\ Standard equilibrium potential of the concerted reaction  as well as the \ce{CoP-}/CoP redox system, relative to RHE (pH = 3).}
  \label{tab:cpet2}
  \begin{center}
  \begin{tabular}{lc}
    \hline
    reaction                                                                                             & $E^\circ$ (V) \\
    \hline
     \ce{CoP} +\ce{e-} $\rightarrow$ \ce{CoP-}                                                           & $-$0.43 \\ 
     \ce{[CoP-CO2]-} + \ce{H2O} +\ce{e-}            $\rightarrow$ \ce{[CoP-COOH]-} + \ce{OH-}            & $-$0.67 \\ 
     \ce{[CoP-CO2]-}\ce{(H2O)2} + \ce{H2O} +\ce{e-} $\rightarrow$ \ce{[CoP-COOH]-}\ce{(H2O)2} + \ce{OH-} & $-$0.80 \\ 
    \hline
  \end{tabular}
  \end{center}
\end{table}

\section{CONCLUSIONS}
We investigated the details of the Koper's water facilitated \ce{CO2} reduction mechanism~\cite{shen2015, gottle2017} using the newly proposed simple but accurate protocol based on thermodynamics, which utilizes DFT calculations and experimental data. 
In our protocol, the three major errors, that from the one-electron reduction of CoP, the hydrogen bonds at \ce{CO2} or \ce{COOH} moiety in \ce{[CoP-CO2]-} and  \ce{[CoP-COOH]-}, and the free energy difference between \ce{H2O} and \ce{OH-}, can be reduced substantially. 
The errors coming from the one-electron reduction potential of CoP and the treatment of the hydrogen bonds are estimated to be the order of 0.01 eV.
From the comparison of the three different pathways, we concluded that the water facilitated \ce{CO2} reduction takes place through the concerted pathway.  
We successfully explained, for the first time, the details of the cobalt-porphyrin catalyzed electrochemical reduction of \ce{CO2} at low pH region (pH of around 3).

The importance of our findings from the point of view of materials design is that the equilibrium potential of Eq.~\ref{eq:concerted} can be controlled by changing the catalyst, meaning it becomes useful guide to design not only the cobalt-porphyrin complexes but also other catalysts, such as MOFs and metal alloys.
So far, the binding energies of H and COOH to the catalyst are used as the guide to design new catalysts.~\cite{cheng2015tailoring, wannakao2017}
However, it is difficult to find such catalysts since usually the binding of \ce{H+} to the negatively charged catalyst is much stronger than that of \ce{CO2} as shown in Ref.~\citenum{cheng2015tailoring}. 
With our new guide, i.e., the redox potential of Eq.~\ref{eq:concerted} (or Eq.~\ref{eq:H2OOH}) and its evaluation protocol, we can expect to find broader candidates due to much milder criteria. 


\begin{suppinfo}
%
%
The following files are available free of charge.
\begin{itemize}
   \item SI\_CoP\_reduction.pdf: Binding energy between \ce{CoP-} and \ce{H+}; geometries and one-electron reduction potentials of CoTPP and CoPc; pKa values of CoP-COOH and [CoP-COOH]$^-$
   \item geometry.txt: Molecular geometries (xyz format) used in the paper.
\end{itemize}
\end{suppinfo}

\providecommand{\latin}[1]{#1}
\makeatletter
\providecommand{\doi}
  {\begingroup\let\do\@makeother\dospecials
  \catcode`\{=1 \catcode`\}=2\doi@aux}
\providecommand{\doi@aux}[1]{\endgroup\texttt{#1}}
\makeatother
\providecommand*\mcitethebibliography{\thebibliography}
\csname @ifundefined\endcsname{endmcitethebibliography}
  {\let\endmcitethebibliography\endthebibliography}{}


\end{document}